\documentclass[prl,twocolumn,aps,showpacs]{revtex4}
\usepackage{graphicx,bm,hyperref,amsmath,amssymb}
\newcommand{\mscr}[1]{{\mbox{\scriptsize #1}}}
\begin{document}
\title{Coherent oscillations of current due to nuclear spins}
\author{Sigurdur I.\ Erlingsson, Oleg N.\ Jouravlev, and Yuli V.\ Nazarov}
\affiliation{Delft University of Technology, Department of NanoScience,
             Lorentzweg 1, 2628 CJ Delft, The Netherlands}
\date{\today}
\begin{abstract}
We propose a mechanism for very slow coherent oscillations 
of current and nuclear spins in a quantum dot system, 
that may qualitatively explain some recent experimental
observations.  
We concentrate on an experimentally relevant double dot
setup where hyperfine interaction lifts the spin blockade. 
We study the dependence of the magnitude
and period of the oscillations on magnetic field and anisotropy.
 
\end{abstract}
\maketitle

There are significant experimental 
and theoretical  efforts aimed to utilize and manipulate the electron
spin in the context of electronic transport, those are commonly
referred to as spintronics\cite{awschalom02:xx,wolf01:1488}. 
Some of them involve nuclear spins as well.
Many efforts concern  GaAs semiconductor structures where the 
hyperfine interaction between electron and nuclear spin
is relatively strong \cite{meier84:381}.
Furthermore, the nuclear spin relaxation times are 
much longer than the time scales related to electron
dynamics\cite{paget77:5780,paget82:4444}.  This time scale separation has
facilitated experiments where a quasi-stationary
polarization of the nuclear system was achieved and its effect on the
electronic transport was observed
\cite{salis01:2677,salis01:195304,wald94:1011}. 
Recent experiments implement novel ways of controlling
the interaction of electron and nuclear spin
\cite{smet02:281,kawakami01:131}, whereby the coherent oscillations 
between 'up' and 'down' polarized nuclear system have been observed
\cite{machida03:409}.    

Transport experiments with quantum dots 
allow for a detailed study and control 
of the quantum dot energy spectrum, both in
the regime of linear transport
and in non-linear regime of excitation
spectroscopy\cite{kouwenhoven97:105,kouwenhoven01:701}.  
They reveal interesting regimes where unusual mechanisms of  
electron transport are the dominant ones. 
For instance, in the Coulomb blockade regime
the direct charging of
the quantum dot is forbidden by energy conservation
and the residual current is due to cotunneling
\cite{averin90:2446,defranceschi01:878}. 
Another regime is the so-called  {\it spin blockade} where 
the direct electron transfers are blocked by virtue of spin
conservation \cite{weinman95:984}.
Spin blockade may be achieved 
in various ways, e.\ g.\ with spin polarized
leads\cite{ciorga02:2177}.  Recent experiment realizes the spin
blockade in  a double dot system, where the absence of transitions 
between spin-singlet and spin-triplet states in the dots  
explains the current rectification observed \cite{ono02:1313}.
Any spin-flip mechanism facilitates these transitions,
giving rise to a small residual current.

The same group has recently reported \cite{ono02:1313b} 
an unexpected and unusual result.
They have observed coherent oscillations 
of the residual current in this regime with a period
in the range of seconds.
This extremely long time scale together with 
the fact that the oscillation period
and amplitude can be modified by
resonant excitation of the nuclear spins,
strongly suggests
that the origin of the oscillations may be traced to
the hyperfine interaction \cite{ono02:1313b}.

In this letter we propose a concrete mechanism for these
oscillations that can at least qualitatively explain
the observation made in  Ref.\ \onlinecite{ono02:1313b}.
The effect comes about from the dynamics of nuclear spins
driven by hyperfine interaction with electron spin. The
configuration of the nuclear spins can be presented
with two effective magnetic fields acting on the electron
spin in the two dots. The difference
of these fields lifts the spin blockade thereby
affecting the average current and electron spin.
These fields are quasistationary at the scale of
successive electron transfers. They
precess around the external magnetic field ($z$-axis)
with frequency $10^7$\,Hz. Albeit this precession
does not manifest itself in current oscillations.
The oscillations result from slow {\it nutations} of the fields.
These nutations arise from small deflection of electron spin from the
$z$-axis, the deflection being induced by the fields. 

The dynamics of the nuclear spin fields appears to be far
from chaotic so that the period and magnitude of
the oscillations strongly depend on initial conditions
of the nuclear spin system. Therefore the comparison
with experiment may be only qualitative. In reality, the
relaxation of nuclear spin  would
lead to stabilization of the oscillations with a certain
amplitude and period. However, such stabilizing mechanisms
would manifest itself at time scale much longer than the period.
This is why we do not consider them in the present model.  

\begin{figure}[ht]
\includegraphics[angle=0,width=7cm]{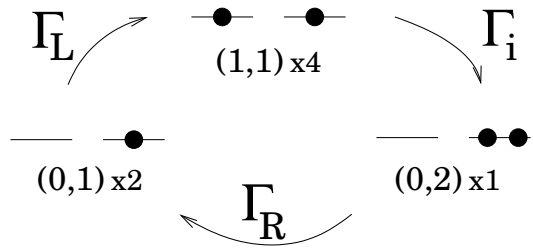}
\caption{Each charge configuration is characterized by ($N_L$,$N_R$) and also
its multiplicity.  
The three charge transfer processes are ($i$) electron entering the
left dot from the left lead, ($ii$) electron going from left dot to
right dot, and ($iii$) electron leaving the right dot to the right lead.
See text for discussion on the $\Gamma$'s.}
\label{fig:fig1}
\end{figure}

For details of the setup we refer the reader to Ref.\
\onlinecite{ono02:1313}. In the regime of interest, the
double dot can be in three distinct charge configurations
as shown in  Fig.\ \ref{fig:fig1}.
A charge configuration is characterized by
($N_L$,$N_R$), $N_L$($N_R$) being the number of electrons in the
left (right) dot. Transitions from (0,1) to (1,1)
and from (0,2) to (0,1) are relatively fast
involving electron tunneling
either from the left or to the right lead with the rates
$\Gamma_{L,R}$. The bottleneck
of the transport cycle are transitions between
(1,1) and (0,2).
In the (0,2) configuration both electrons share
the same orbital state, this makes it
a non-degenerate spin singlet.
In contrast to this, the (1,1) configuration
comprises 4 possible states, grouped into spin-singlet
and spin-triplet. 
The transitions between
the (1,1) singlet and (0,2) occur with the rate
$\Gamma_i$ that is determined by the
tunneling amplitude between the  dots and a relevant 
mechanism of inelastic scattering, e.g.\ phonons.
These transitions do not require any spin-flip.
The transitions between the (1,1) triplet states
and (0,2) singlet are forbidden by spin conservation:
this causes the spin blockade. 

The hyperfine interaction with nuclei induces mixing of 
the singlet and triplet states in the $(1,1)$
configuration. The part of the total Hamiltonian
which contains the
electron and nuclear spin operators reads 
\begin{eqnarray}
H_s= \hat{\bm{K}_L} \cdot \hat{\bm{S}}_L+
\bm{K}_R \cdot \hat{\bm{S}}_R +E_Z\cdot
(\hat{S}^z_L+\hat{S}^z_R) + \nonumber \\
\Delta_{ST}
(\hat{\bm{S}}_L+\hat{\bm{S}}_R)^2/2.
\label{eq:spinHamiltonian}
\end{eqnarray}
Here $\hat{\bm{S}}_{L,R}$ are operators of electron
spin in each dot
\footnote{The spin operator are $\hat{\bm{S}}_\alpha =\sum_{\eta,\gamma}
\hat{\bm{\sigma}}_{\eta \gamma} d_{\alpha\eta}^{\dagger} d_{\alpha \eta}$ where
$\hat{\bm{\sigma}}=(\hat{\sigma}_x,\hat{\sigma}_y,\hat{\sigma}_z)$
are the Pauli matrices and $d_{L\eta}^{\dagger}$ creates an electron
with spin $\eta$ in dot $\alpha$.}. 
The effect of nuclear spins is combined into two effective fields
$\hat{\bm{K}}_{L,R}$. In each dot
\begin{equation}
\hat{\bm{K}}_{L,R}=\hbar
\sum_{k} \gamma_{L,R;k} \hat{\bm{I}}_k; \;\; \gamma_{L,R;k} = \hbar^{-1}A
|\psi_{L,R}(\bm{R}_k)|^2  
\end{equation}
where $\hat{\bm I}_k$ being operators of nuclear spin at $\bm{R}_k$,
the summation goes over all nuclei and precession frequencies
$\gamma_k$ are set by an envelop of electron wave function and
hyperfine constant $A$. 
The third term in Eq.\ \ref{eq:spinHamiltonian} represents the Zeeman
splitting $E_Z$ in the external magnetic field while the fourth term
represents the exchange splitting between the singlet and triplet. 
We adopt  a semiclassical approximation of the effective fields
$\bm{K}_{L,R}$ replacing them by classical
variables\cite{dyakonov86:110,merkulov02:205309,erlingsson02:155327}.  
This approximation is justified by a big number of nuclei in the dots, 
$N_\mscr{QD} \gg 1$. The third  and fourth terms in  Eq.\
(\ref{eq:spinHamiltonian}) include the full spin only and therefore
split the states onto singlet and three Zeeman-split triplet
components. The mixing of these states is proportional to the
difference of two effective fields $\bm{K}_A\equiv\bm{K}_L-\bm{K}_R$.   

The mixing thus lifts spin blockade.
We assume that this mixing is the only mechanism
of the residual current.
This assumption is not crucial since
alternative mechanisms, those include
co-tunneling and non-nuclear spin-flips,
would only produce an extra d.c.\ current background
for nuclear-induced current oscillations. 

We solve the problem in two steps.
First, we solve for the density matrix of electron
states assuming stationary $\bm{K}_{L,R}$.
The output of the calculation are the average
current and the average electron spin $\langle \bm{S}_{L,R} \rangle$  
in terms of $\bm{K}_{L,R}$.
Second, we use this output to derive equations
for the dynamics of $\bm{K}_{L,R}$ and subsequently
analyze this dynamics.
This approach relies on the time scale separation:
the fields $\bm{K}_{L,R}$ should not change
at the time scale of successive electron transfers.
The transfer rate can be estimated as ($E_Z \simeq \Delta_{ST}$)
$(K_A/\Delta_{ST})^2\Gamma_i$.
The small factor
$(K_A/\Delta_{ST})$ is the ratio of the mixing
and energy difference between singlet and triplet
states and quantifies
suppression of the current in the spin blockade regime.
The fastest nuclear spin
motion  that changes $\bm{K}$
is the precession around external magnetic field 
with frequency $\omega_{NMR} \simeq 10^7$\,Hz. 
This results in the condition 
$\Gamma_i \gg (\Delta_{ST}/K_A)^2 \omega_{NMR}$
for the validity of our approach.
As we see below, the current oscillations
are much slower with a typical period of the order of
$\gamma^{-1}(\Delta_{ST}/K_A)$, where
$\gamma=(\gamma_L+\gamma_R)/2$. The precession frequency 
and typical magnitude of effective field
can be estimated \cite{erlingsson02:155327} as $\gamma \simeq E_n /N_{QD}$,
$K \simeq E_n /\sqrt{N_{QD}}$, where $E_n\approx0.135$\,meV
in GaAs and $N_\mscr{QD}\approx$10$^6$ for
the quantum dots in question. The exchange splitting
$\Delta_{ST} \propto 10^{-5}-10^{-4} eV$ as estimated
in Ref.\ \onlinecite{ono02:1313b}. This gives
$\gamma^{-1}(\Delta_{ST}/K_A) \simeq 0.01-0.1$\,sec. 

To make the first step, we
describe the evolution of the electron system with
Bloch equations for the density matrix approach.  
There are seven quantum states involved
in the transport ($|s \rangle$ for the singlet in the (0,2)
configuration, $|+ \rangle$ and
$|- \rangle$ for the two doublet components
in the (1,0) configuration
,$|0 \rangle$ for the singlet and $|1 \rangle,|2 \rangle,|3 \rangle$ for
triplet states 
in the (1,1) configuration, those correspond to $S_z=1,0,-1$
respectively) so that the full density matrix
comprises of $49$ elements. However, we can disregard
most of the non-diagonal elements of the matrix
except those between 4 states of the (1,1) configuration.
So we end up with 19 equations only.
We present here 3, this suffices to
illustrate the overall structure.
\begin{subequations}
\label{eq:rho}
\begin{eqnarray}
\frac{d\rho_{11}}{dt}&=& \frac{\Gamma_L}{2}\rho_{++}
                +\Im \{ K^+_A \rho_{10}\}  
\label{eq:rho11}\\
\frac{d\rho_{00}}{dt}&=&-\Gamma_i \rho_{00}+\frac{\Gamma_L}{4}\bigl
(\rho_{++}+\rho_{--}\bigr ) 
\nonumber \\
               & & +\Im \{-K_A^+\rho_{10}+K_{A}^z\rho_{20}+K_A^- \rho_{30}
                \} \label{eq:rho00}\\
\frac{d\rho_{10}}{dt}&=&-i\bigl (\Delta_{ST} + E_Z)
+i\frac{\Gamma_i}{2}\bigr )\rho_{10} 
+\frac{1}{2}\bigl (-i K_A^-( \rho_{11}\!\!-\!\!\rho_{00})
                  \nonumber  \\
               & & +iK_{A}^z \rho_{12} 
                  +i K_A^+ \rho_{13}\bigr ) \label{eq:rho10},
\end{eqnarray}
\end{subequations}
where $K_A^\pm=(K_{A}^x \pm iK_{A}^y)/\sqrt{2}$.
Note that the inelastic rate $\Gamma_i$ does not
appear in Eq.\ (\ref{eq:rho11}), the same is true for the other
triplets, but it is present in Eq.\ (\ref{eq:rho00}) for the
singlet.  
The average electron spins in each dot and the
current can be readily obtained from the
 stationary solution $\hat{\rho}^\mscr{\,st}$
 of Eq.\ ({\ref{eq:rho}}):
$\langle{\bm{S}}_{L,R}\rangle=
\mbox{Tr}\{\hat{\rho}^\mscr{\,st} \hat{\bm{S}}_{L/R}\}$,
$I=e \Gamma_{R} \rho^\mscr{\,st}_{ss}$.
We do not need to present the cumbersome
full solutions for the average spin
here. Instead, we assume $\Gamma_{in} \ll
\Gamma_{L,R},\Delta_{ST}/\hbar$ 
and note that in the zeroth order in $K_A/\Delta_{ST}$
the average spin is in the $z$-direction and
($x_B \equiv E_Z/\Delta_{ST}$)
\begin{equation}
\langle S^z_{L,R} \rangle=-\frac{2 E_Z}{\Delta_{ST}} S;
\;\; S^{-1} \equiv 2(1+x^2_B)+
\left(\frac{|K_A^+|}{K_{Az}}\right)^2.
\end{equation}
With the same accuracy the current reads
\begin{equation}
\frac{I}{e}=\left(\frac{|K_A^+|}{\Delta_{ST}}\right)^2
4 S\Gamma_i.
\label{eq:current}
\end{equation}
In addition, we need the deflections of $\langle \bm{S}_{L,R}\rangle$
from the $z$-direction. They arise in the next order
in $K_A/\Delta_{ST}$ and are antiparallel in opposite dots:
\begin{equation}
\left(\begin{array}{ccc} \langle S^x_{L,R} \rangle  \cr\langle
S^y_{L,R} \rangle   \cr
\langle S^z_{L,R} \rangle \end{array}\right)= \pm \frac{S}{2 \Delta_{ST}}
\left(\begin{array}{ccc} K^x_A \cr K^y_{A} \cr
\frac{|K^+_A|^2}{K^z_A}\end{array}\right).
\end{equation}

Now we perform the second step and describe the dynamics of the
effective fields. To simplify, we will assume the
same electron precession frequencies for all nuclei in
each dot: $\gamma_{L,R;k}=\gamma_{L,R}$. The advantage
of this model is a closed system of equations
for the collective fields:
\begin{eqnarray}
\frac{d}{dt}\bm{K}_\alpha=\gamma_\alpha \langle \bm{S}_\alpha \rangle
\times \bm{K}_\alpha 
               + \gamma_\mscr{GaAs} \bm{B}\times  \bm{K}_\alpha
\label{eq:precession}
\end{eqnarray}
($\alpha=L,R$) that describes precession of these
fields, their moduli $K_{L,R}$ being constants of motion.
The main precession is around external magnetic field
with the frequency $\gamma_\mscr{GaAs} |B|$.
However, this precession is irrelevant since
it does not change the $K^{z}_{L,R}$ and
the angle $\phi$ between the projections of these
vectors onto $xy$ plane. These three
slow variables actually determine
the average spin and current, and the evolution
equations for those
\begin{eqnarray}
\frac{d}{dt}K^z_\alpha&=& \gamma_\alpha ( \langle S^x_\alpha\rangle
 K^x_\alpha - \langle S^y_\alpha \rangle K^y_\alpha)\nonumber \\
\frac{d}{dt}\phi &=& \gamma_L \langle S^z_L \rangle -\gamma_R \langle
 S^z_R\rangle ,
\label{eq:evolution}
\end{eqnarray}
do not contain
$\gamma_\mscr{GaAs} B \gg \gamma_{L,R} \langle S_{L,R}\rangle $.
These three equations have an extra integral
of motion, 
$K^z_S \equiv (K^z_L \gamma_R + K^z_R
\gamma_L)/2\gamma$.
Moreover, the equation for 
two remaining variables appears to be of a Hamiltonian type.
In dimensionless variables $k \equiv  K^z_{A}/K_L, \tau = t \gamma
K_L/\Delta_{ST}$ the equations read
\begin{equation}
\frac{d}{d\tau}
\left(\begin{array}{cc}
 k \cr  
\phi \end{array}\right) = \frac{k^2}{(b - \cos\phi)Y}
\left(\begin{array}{rr}\frac{\partial}{\partial \phi}\cr
 -\frac{\partial }{\partial k} \end{array}\right) {\cal L}
\end{equation}
where 
\begin{eqnarray}
\mathcal{L}(k,\phi)&=&X(k)+Y(k)\cos\phi,\nonumber\\
X(k)&=&-\frac{1+k_R^2-2(k^z_{S})^2}{2} k +2 \epsilon x_B
k^2+\frac{k^3}{4},\nonumber\\
Y(k)&=&k\sqrt{(1-(k^z_{S}+k/2)^2)
(k_R^2-(k^z_{S}-k/2)^2)},
\nonumber \\
b(k)&=&\frac{
(\frac{7}{4}+2x_B^2)k^3+\frac{k}{2}(1+k_R^2-2(k^z_{S})^2)
}{Y(k)}, 
\end{eqnarray}
and we introduced the  notations $k_R,k^z_S \equiv
K_R/K_L,K^z_S/K_L$ and
$\epsilon = (\gamma_L -\gamma_R)\Delta_{ST}/2\gamma K_L$.
We also assume here
that the asymmetry of electron precession frequencies is small,
$\delta \gamma \equiv|\gamma_L-\gamma_R|\ll \gamma_{L,R}$.
One would expect this for the experiment in question 
since the two dots are nominally identical

\begin{figure}[ht]
\includegraphics[angle=-90,width=8cm]{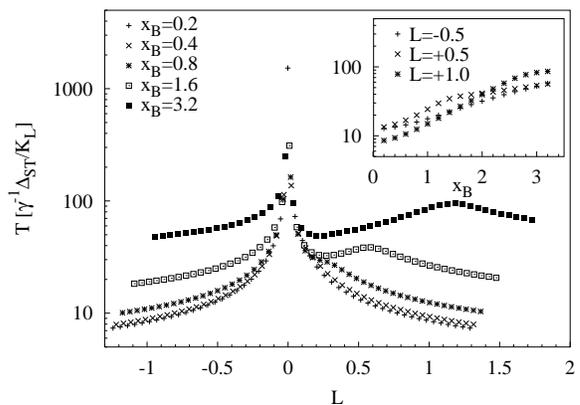}
\caption{The period as a function the integral of motion $L$ for
various values of $x_B$.  The asymmetry is $\epsilon=0.1$, $k_R=0.9$
and $k_S^z=0.0$. The inset shows the period as a function of $x_B$ for
fixed values of $L$.}
\label{fig:fig2}
\end{figure}

The $\cal{L}$ is evidently yet another constant of motion that
depend on initial condition of the nuclear system.
The solution of $\mathcal{L}=L$, if it exists, determines a closed 
orbit in the $(k,\phi)$ phase space and the system 
experiences periodic 
motion along this orbit. This motion manifests itself in 
the current oscillations by virtue of Eq.\ \ref{eq:current}.
The period and magnitude of the oscillations do depend on the initial
conditions $L,k_R,k^z_S$. If $L$ is close to $0$, 
the period even diverges.
There is, however, some regular dependence
on the asymmetry $\epsilon$ and the external magnetic field that
enters through parameter $x_B$, see Fig.\ \ref{fig:fig2}. This
dependence can be summarized as follows. 
The period $T$ depends on asymmetry. If $\epsilon x_B \lesssim 1$,  
$T \simeq \gamma^{-1} \Delta_{ST}/K_{L}$
for $x_B \lesssim 1$ and $T \simeq \gamma^{-1} E^2_Z/\Delta_{ST} K_L$
for $x_B \gg 1$. 
In the opposite case of relatively
large asymmetry $ \epsilon x_B \gg 1$, $T \simeq (x_B \delta \gamma)^{-1}$
for $x_B \ll 1$ and $T \simeq x_B (\delta \gamma)^{-1}$ for $x_B \gg 1$.
Also, the amplitude of the oscillations relative
to the average current is of the order of $1$ for $x_B \lesssim 1$
and decreases as $x^{-2}_B$ for $x_B \gg 1$. 

\begin{figure}[ht]
\includegraphics[angle=-90,width=8cm]{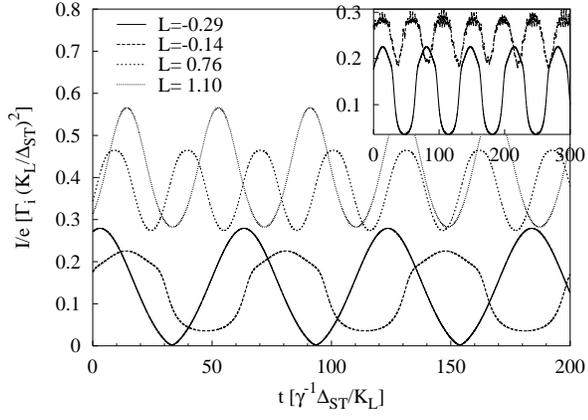}
\caption{The current shown as a function of time for $\epsilon$=0.1,
$x_B$=1.6, but for different initial conditions, i.e.\ different $L$.
The inset shows current for $L=-0.14$ in the case of $N_b$=1 and
25, where $N_b$ is the number of blocks.  Note that the long period
oscillations are still present for $N_b \gg 1$.} 
\label{fig:fig3}
\end{figure}

The initial values of $\bm{K}_\alpha$ that determine the 
actual magnitude and period of the oscillations are distributed
according to Gaussian statistics \cite{erlingsson02:155327}
with average squares corresponding to average squares of total
nuclear spins in the dot. In Fig.\ \ref{fig:fig3} the current, see Eq.\
(\ref{eq:current}), is plotted for various initial conditions but
fixed $\epsilon=0.1$ and $x_B=1.6$. We note apparent anharmonicity of
the oscillations, this feature has been stressed in Ref.\
\onlinecite{ono02:1313b}. 

One might think that the periodic oscillations we obtained in
our approach is an artefact of oversimplified model for nuclear
dymanics in use. Recent work emphasizes the importance of
the fact that precession frequency $\gamma_k$ varies 
from nucleus to nucleus \cite{khaetskii02:186802}. This issue can be
addressed within the semiclassical approach used here. To implement
such approach numerically, 
we separate the nuclear spin system into $N_b \gg 1$ blocks where
the $\gamma_k$'s are the same 
within each block but differ from block to block.
The number of spins remains large so that the nuclear dynamics
can be described in terms of effective fields $\bm{K}^{(b)}_{L,R}$.   
This results in $6 N_b$ evolution equations similar to Eqs. 
\ref{eq:evolution}. Intuitively, one expects such complicated
dynamics to be chaotic, so that the memory about initial conditions
is lost after some time  $\simeq \gamma^{-1}$. 
This would be really dreadful for the mechanism discussed, so we have
performed extensive numerical simulations to check this circumstance 
\cite{erlingsson03:xx}. 
To summarize the results, the dynamics is not chaotic, the memory
about initial conditions persists and the nuclear system exhibits
regular oscillations, typically with several periods. To illustrate 
this fact, we present a typical result in the inset of Fig. 3.
It shows the regular long-period motion and extra fast oscillations  
on the timescale of $\gamma^{-1}$.  

In conclusion, we propose a mechanism whereby the transport
via a double quantum dot induces slow regular nutations
of the nuclear spin system, these nutations are seen in the
transport current. We model the concrete experimental situation 
(\cite{ono02:1313b}) and our estimations of the typical frequency, 
anharmonic shape of the oscillation predicted, and 
sensitivity to magnetic field shown correspond to the observations
made in  \cite{ono02:1313b}. More research on relevant nuclear
spin relaxation mechanisms is needed for detailed comparison
with the experiment. From the other hand, the mechanism presented
is sufficiently general and works in any conditions where 
the hyperfine interaction provides the main mechanism of 
spin blockade lifting.

We are grateful to the authors of Ref. \cite{ono02:1313b}
for communicating  their results prior to publication.  
We acknowledge  the financial support by FOM.  
\end{document}